\documentstyle[emulateapj5,epsf]{aastex}
\def \farcs{\hbox{$.\!\!^{\prime\prime}$}}
\def \farcm{\hbox{$.\!\!^{\prime}$}}

\def \lumstar{$L_B=10^{10}h^{-2} {\rm L}_{\rm {B}\odot}~$}
\def \lumhoek{$5.6 \times 10^{9}h^{-2} {\rm L}_{\rm {B}\odot}~$}
\def \lumstarz{$L_B(z=0)=10^{10}h^{-2} {\rm L}_{\rm {B}\odot}~$}

\lefthead{Hoekstra et al.}
\righthead{Properties of dark matter halos}
\begin{document}

\title{Properties of galaxy dark matter halos from weak lensing}

\author{Henk~Hoekstra$^{1,2,3}$, H.K.C.~Yee$^{2,3}$, and Michael
D.~Gladders$^{3,4}$}

\begin{abstract}

We present the results of a study of weak lensing by galaxies based on
45.5 deg$^2$ of $R_C$ band imaging data from the Red-Sequence Cluster
Survey (RCS). We define a sample of lenses with $19.5<R_C<21$, and a
sample of background galaxies with $21.5<R_C<24$. 

We present the first weak lensing detection of the flattening of
galaxy dark matter halos. We use a simple model in which the
ellipticity of the halo is $f$ times the observed ellipticity of the
lens. We find a best fit value of $f=0.77^{+0.18}_{-0.21}$, suggesting
that the dark matter halos are somewhat rounder than the light
distribution. The fact that we detect a significant flattening implies
that the halos are well aligned with the light distribution. Given the
average ellipticity of the lenses, this implies a halo ellipticity of
$\langle e_{\rm halo} \rangle=0.33^{+0.07}_{-0.09}$, in fair agreement
with results from numerical simulations of CDM. We note that this
result is formally a lower limit to the flattening, since the
measurements imply a larger flattening if the halos are not aligned
with the light distribution. Alternative theories of gravity (without
dark matter) predict an isotropic lensing signal, which is excluded
with 99.5\% confidence. Hence, our results provide strong support for
the existence of dark matter.

We also study the average mass profile around the lenses, using a
maximum likelihood analysis. We consider two models for the halo mass
profile: a truncated isothermal sphere (TIS) and an NFW profile. We
adopt observationally motivated scaling relations between the lens
luminosity and the velocity dispersion and the extent of the halo.
The TIS model yields a best fit velocity dispersion of
$\sigma=136\pm5\pm3$ km/s (all errors are 68\% confidence limits; the
first error bar indicates the statistical uncertainty, whereas the
second error bar indicates the systematic error) and a truncation
radius $s=185^{+30}_{-28} h^{-1}$ kpc for a galaxy with a fiducial
luminosity of \lumstar (under the assumption that the luminosity does
not evolve with redshift). Alternatively, the best fit NFW model
yields a mass $M_{200}=(8.4\pm0.7\pm0.4)\times 10^{11} h^{-1} M_\odot$
and a scale radius $r_s=16.2^{+3.6}_{-2.9} h^{-1}$ kpc. This value for
the scale radius is in excellent agreement with predictions from
numerical simulations for a halo of this mass.

\end{abstract}


\section{Introduction}

The existence of massive dark matter halos around galaxies is widely
accepted, based on different lines of evidence, such as flat rotation
curves of spiral galaxies (e.g., Van Albada \& Sancisi 1986) and
strong lensing systems (e.g., Keeton, Kochanek \& Falco
1998). However, relatively little is known about the properties of
dark matter halos. Strong lensing only probes the gravitational
potential on small (projected) scales, whereas the lack of visible
tracers at large radii hamper dynamical methods. To date, only
satellite galaxies have provided some information (e.g., Zaritsky \&
White 1994; McKay et al. 2002; Prada et al. 2003).

A promising approach to study the galaxy dark matter halos is weak
gravitational lensing. The tidal gravitational field of the dark
matter halo introduces small coherent distortions in the images of
distant background galaxies. The weak lensing signal can be measured
out to large projected distances~~from~~the~~lens,~~and hence~~provides 
a~~unique 

\vbox{
\vspace{0.5cm}
\footnotesize
\noindent 
$^1$~CITA, University of Toronto, Toronto, Ontario M5S 3H8, Canada\\
$^2$~Department of Astronomy, University of Toronto,
Toronto, Ontario M5S 3H8, Canada\\
$^3$~Visiting Astronomer, Canada-France-Hawaii Telescope, which
is operated by the National Research Council of Canada, Le Centre 
National de Recherche Scientifique, and the University of Hawaii\\
$^4$~Observatories of the Carnegie Institution of 
Washington, 813 Santa Barbara Street, Pasadena, California 91101}

\noindent probe of the gravitational potential on large scales.

The applications of this approach are numerous: one can infer masses
of galaxies and compare the results to their luminosities (e.g., McKay
et al. 2001; Wilson et al. 2001), or one can attempt to constrain the
halo mass profile (e.g., Brainerd et al. 1996; Hudson et al. 1998;
Fischer et al. 2000; Hoekstra et al. 2003). Also, weak lensing can be
used to constrain the shapes of halos by measuring the azimuthal
variation of the lensing signal. Unfortunately, one can only study
ensemble averaged properties, because the weak lensing signal induced
by an individual galaxy is too low to be detected.

A successful measurement of the lensing signal requires large samples
of both lenses and background galaxies.  The first attempt to detect
the lensing signal by galaxies was made by Tyson et al. (1984) using
photographic plates. It took more than a decade and CCD cameras before
the first detections were reported (Brainerd et al. 1996; Griffiths et
al. 1996; Dell'Antonio \& Tyson 1996; Hudson et al. 1998). These early
results were limited by the small areas covered by the observations.

The accuracy with which the galaxy-galaxy lensing signal can be
measured depends on the area of sky that is observed, and on the
availabitity of redshifts for the lenses (as it allows for a proper
scaling of the lensing signal). Photometric redshifts were used by
Hudson et al. (1998) to scale the lensing signal of galaxies in the
Hubble Deep Field, and by Wilson et al. (2001) who measured the
lensing signal around early type galaxies as a function of redshift.
Furthermore, several lensing studies targeted regions covered by
redshift surveys.  Smith et al. (2001) used 790 lenses from the Las
Campanas Redshift Survey; Hoekstra et al. (2003) used 1125 lenses from
the Canadian Network for Observational Cosmology Field Galaxy Redshift
Survey (CNOC2). The areas covered by these surveys are relatively
small.

The Sloan Digital Sky Survey (SDSS) combines both survey area and
redshift information. Its usefulness for galaxy-galaxy lensing was
demonstrated clearly by Fischer et al. (2000). More recently, McKay et
al. (2001) used the available SDSS redshift information to study the
galaxy-galaxy lensing signal as a function of galaxy properties (also
see Guzik \& Seljak 2002; Seljak 2002).

The data used in this paper currently lacks redshift information for
the lenses. However, compared to previous work, the combination of
large area and depth of our observations allow us to measure the
galaxy lensing signal with great precision. We use 45.5 deg$^2$ of
$R_C$-band imaging data from the Red-Sequence Cluster Survey (RCS).
These data have been used previously for several weak lensing
studies. Hoekstra et al. (2002a; 2002b) placed joint constraints on
$\Omega_m$ and $\sigma_8$ by measuring the lensing signal caused by
large scale structure.  Related to the subject of this paper is the
study of the bias parameters as a function of scale by Hoekstra et
al. (2001b) and Hoekstra et al. (2002c). The latter studies made use
of the galaxy-mass cross-correlation function measured from the RCS
data.  Here we use the galaxy-mass cross-correlation function for a
different purpose: we effectively deconvolve the cross-correlation
function to study the properties of dark matter halos surrounding
galaxies at intermediate redshifts.

The structure of the paper is as follows. In \S2 we briefly discuss
the data and the redshift distributions of the lenses and the sources.
The ensemble averaged tangential shear around the lenses (galaxy-mass
cross-correlation function) is presented in \S3. In \S4 we use a
maximum likelihood analysis to derive constraints on the extent of
dark matter halos. The measurement of the projected shapes of the
halos is presented in \S5.

\section{Observations and analysis}

We use the $R_C$-band imaging data from the Red-Sequence Cluster
Survey (Yee \& Gladders 2001; Gladders \& Yee 2003). The complete
survey covers 90 deg$^2$ in both $R_C$ and $z'$, spread over 22 widely
separated patches of $\sim 2.1\times 2.3$ degrees. In this paper we
use data from the northern half of the survey, which consists of 10
patches, observed with the CFH12k camera on the CFHT. These data cover
45.5 deg$^2$ on the sky, but because of masking the effective area is
somewhat smaller. In the lensing analysis we use a total of 42
deg$^2$.  A detailed description of the data reduction and object
analysis can be found in Hoekstra et al. (2002a), to which we refer
for technical details. Here we present a short description of the
various steps in the analysis.

We use single exposures in our analysis, and consequently cosmic rays
have not been removed. However, cosmic rays are readily eliminated
from the photometric catalogs: small, but very significant objects are
likely to be cosmic rays, or artifacts from the CCD. The object finder
gives fair estimates of the object sizes, and we remove all objects
smaller than the size of the PSF. Some faint cosmic rays may hit
galaxies, and consequently might not be recognized as cosmic
rays. Based on the number of cosmic ray hits, and the area covered by
galaxies we find that less than $0.2\%$ of the galaxies might be
affected. Also, cosmic rays only introduce additional noise in the
shape measurement, but do not bias the result. Consequently we
conclude that remaining cosmic rays have a negligble effect on our
results.

The objects in this cleaned catalog are then analysed, which yield
estimates for the size, apparent magnitude, and shape parameters
(polarisation and polarisabilities). The objects in this catalog are
inspected by eye, in order to remove spurious detections.  These
objects have to be removed because their shape measurements are
affected by cosmetic defects (such as dead columns, bleeding stars,
halos, diffraction spikes) or because the objects are likely to be
part of a resolved galaxy (e.g., HII regions).

To measure the small, lensing induced distortions in the images of the
faint galaxies it is important to accurately correct the shapes for
observational effects, such as PSF anisotropy, seeing and camera
shear; PSF anisotropy can mimic a cosmic shear signal, and a
correction for the seeing is required to relate the measured shapes to
the real lensing signal. To do so, we follow the procedure outlined in
Hoekstra et al. (1998).  We select a sample of moderately bright stars
from our observations, and use these to characterize the PSF
anisotropy and seeing.  We fit a second order polynomial to the shape
parameters of the selected stars for each chip of the CFH12k
camera. These results are used to correct the shapes of the galaxies
for PSF anisotropy and seeing.

The effect of the PSF is not the only observational distortion that
has to be corrected. The optics of the camera stretches the images of
galaxies (i.e., it introduces a shear) because of the non-linear
remapping of the sky onto the CCD.  We have used observations of
astrometric fields to find the mapping between the sky and the CCD
pixel coordinate system, and derived the corresponding camera shear,
which is subsequently subtracted from the galaxy ellipticity (see
Hoekstra et al. 1998).

The findings presented in Hoekstra et al. (2002a) suggest that the
correction for PSF anisotropy has worked well. The absence of a
``B''-mode on large scales in the measurements of the cosmic shear
(Hoekstra et al. 2002b) provides additional evidence that systematics
are well under control (the small scale ``B''-mode is attributed
to intrinsic alignments). Furthermore, cosmic shear studies are much
more sensitive to systematics than galaxy-galaxy lensing measurements
(e.g., see Hoekstra et al. 2003). In galaxy-galaxy lensing one
measures the lensing signal that is perpendicular to the lines
connecting many lens-source pairs. These connecting lines are randomly
oriented with respect to the PSF anisotropy, and hence suppress any
residual systematics.

\subsection{Redshift distributions}

For the analysis presented here, we select a sample of ``lenses'' and
``sources'' on the basis of their apparent $R_C$ magnitude. We define
galaxies with $19.5<R_C<21$ as lenses, and galaxies with $21.5<R_C<24$
as sources which are used to measure the lensing signal. This
selection yields a sample of $\sim 1.2\times 10^5$ lenses and $\sim
1.5\times 10^6$ sources.

For a singular isothermal sphere, the amplitude of the lensing signal
depends on $\langle\beta\rangle$, the average ratio of the angular
diameter distances between the lens and the source, $D_{ls}$, and the
distance between the observer and the source $D_s$. More general, the
signal also depends on $D_l$, the distance between the observer and
the lens. Hence, to interpret the measurements, such as size and mass,
one needs to know the redshift distributions of both lenses and
sources.

The CNOC2 Field Galaxy Redshift Survey (e.g., Lin et al. 1999; Yee et
al. 2000; Carlberg et al. 2001) has measured the redshift distribution
of field galaxies down to $R_C=21.5$, which is ideal, given our limits
of $19.5<R_C<21$. The derived redshift distribution gives a median
redshift $z=0.35$ for the lens sample. In addition, we use the
redshifts and the colors of the galaxies observed in the CNOC2 survey
to compute their rest-frame $B$ luminosity. 

Compared to studies using SDSS data (McKay et al. 2001) we have the
disadvantage that we do not have (spectroscopic) redshifts for the
individual lenses. As shown by Schneider \& Rix (1997) and Hoekstra et
al. (2003) this limits the accuracy of the measurements. We can derive
useful constraints on the masses and extent of dark matter halos, but
we have to assume scaling relations between these parameters and the
luminosity of the galaxies. Multi-color data for the northern part of
the RCS will be available in the near future, allowing us to select a
sample of lenses based on their photometric redshifts, and constrain
the scaling relations.

Nevertheless, the large area covered by the RCS allows us to derive
interesting information about the properties of the lenses. In \S4 we
use different cuts in apparent magnitude to study the properties of
dark matter halos using a maximum likelihood analysis.

For the source galaxies the situation is more complicated.  These
galaxies are generally too faint for spectroscopic surveys, although
recently Cohen et al. (2000) measured spectroscopic redshifts around
the Hubble Deep Field North down to $R_C\sim 24$. Cohen et al. (2000)
find that the spectroscopic redshifts agree well with the photometric
redshifts derived from multi-color photometry. Because of likely
field-to-field variations in the redshift distribution, we prefer to
use the photometric redshift distributions derived from both Hubble
Deep Fields (Fern{\'a}ndez-Soto et al. 1999), which yields a median
redshift of $z=0.53$ for the source galaxies. 

The adopted source and lens redshift distributions result in an
average value of $\langle\beta\rangle=0.29\pm0.01$ (average for the
full sample of lenses and sources), where the error bar is based on
the field-to-field variation and the finite number of galaxies in the
Hubble Deep Fields. The uncertainty in $\langle\beta\rangle$ affects
predominantly our estimates for the galaxy masses (and velocity
dispersions), but is negligible for the other model parameters. We
note that for the galaxy-galaxy lensing analysis presented here the
relevant parameter is $\langle L_B^{0.3}\beta\rangle$, as opposed to
simply $\langle\beta\rangle$, but we have verified that results in a
negligible change in the adopted systematic error. Throughout the
paper we indicate the systematic error in the masses and velocity
dispersions by a second error bar.

\vbox{
\begin{center}
\leavevmode 
\hbox{% 
\epsfxsize=8.5cm 
\epsffile{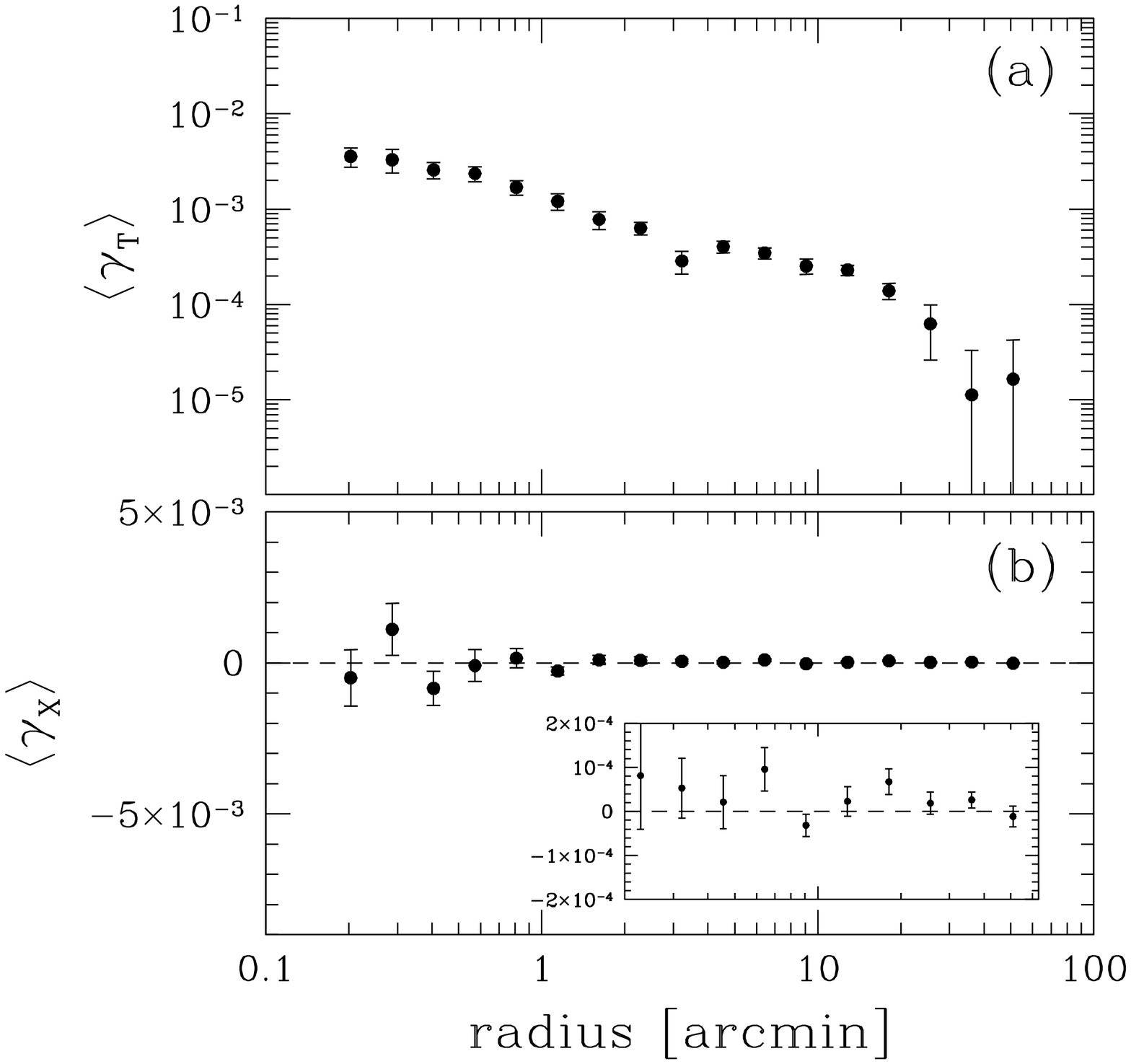}}
\figcaption{\footnotesize (a) The galaxy-mass cross-correlation
function as a function of angular scale. The lenses are selected on
the basis on their apparent $R_C$-band magnitude, taking
$19.5<R_C<21$. The tangential aligment is detected out to a radius of
1 degree. The signal on these large scale no longer reflects the mass
of the lens, but the clustering properties of the lenses. (b) The
signal when the phase of the shear is increased by $\pi/4$. No signal
should be present if the signal detected in panel~(a) is caused by
gravitational lensing. The results are consistent with no signal. The
error bars are so small that we expanded the y-axis by a factor 25 in
the inset of panel~(b).  The complex geometry of the survey on large
scales (holes in the survey area because of masking) results in
non-vanishing values of $\gamma_{\rm X}$. Despite these complications, the
residuals are remarkably small.
\label{gtprofall}}
\end{center}}

\section{Galaxy-mass cross-correlation function}

The galaxy-mass cross-correlation function provides a convenient way
to present the measurements. It is obtained from the data by measuring
the tangential alignment of the source galaxies with respect to the
lens as a function of radius. Its use for studying the halos of
galaxies is limited, because the clustering of galaxies complicates a
direct interpretation of the signal: on small scales the signal is
dominated by the mass distribution of the lens, but on larger scales
one measures the superposition of the contributions from many lenses.

The observed galaxy-mass cross-correlation function as a function of
angular scale is presented in Figure~\ref{gtprofall}a. A significant
signal is detected out to one degree from the lens. If the signal
presented in Figure~\ref{gtprofall}a is caused by gravitational
lensing, no signal should be present when the phase of the distortion
is increased by $\pi/4$ (i.e., when the sources are rotated by 45
degrees). The results of this test, shown in Figure~\ref{gtprofall}b,
suggest that residual systematics are negligible.

Before we can interpret the results we need to examine the
contribution of foreground galaxies. Some of the source galaxies will
be in front of the lenses, and lower the lensing signal independent of
radius. This is absorbed in the value of $\langle\beta\rangle$. These
galaxies decrease the lensing signal independent of angular scale.
Some sources, however, are physically associated with the
lenses. These galaxies cluster around the lenses, affecting the
lensing signal more on small scales. We need to account for this
source of contamination. To do so, we measure a fractional excess of
sources around lenses which decreases with radius as $f_{\rm
bg}(r)=0.93 r^{-0.76}$ (r in arcseconds), similar to what was found by
Fischer et al. (2000).  Under the assumption that the orientations of
these galaxies are random (the tidal interaction with the lens has not
introduced an additional tangential or radial alignment), the observed
lensing signal has to be increased by a factor $1+f_{\rm bg}(r)$. This
assumption is supported by the findings of Bernstein \& Norberg (2002)
who examined the tangential alignment of satellite galaxies around
galaxies, extracted from the 2dF Galaxy Redshift Survey. The
measurements presented in Figures~\ref{gtprofall} have been corrected
for this decrease in signal. We note that the correction for the
presence of satellite galaxies is small, and has no significant effect
on our results.

\vbox{
\begin{center}
\leavevmode 
\hbox{% 
\epsfxsize=8.5cm 
\epsffile{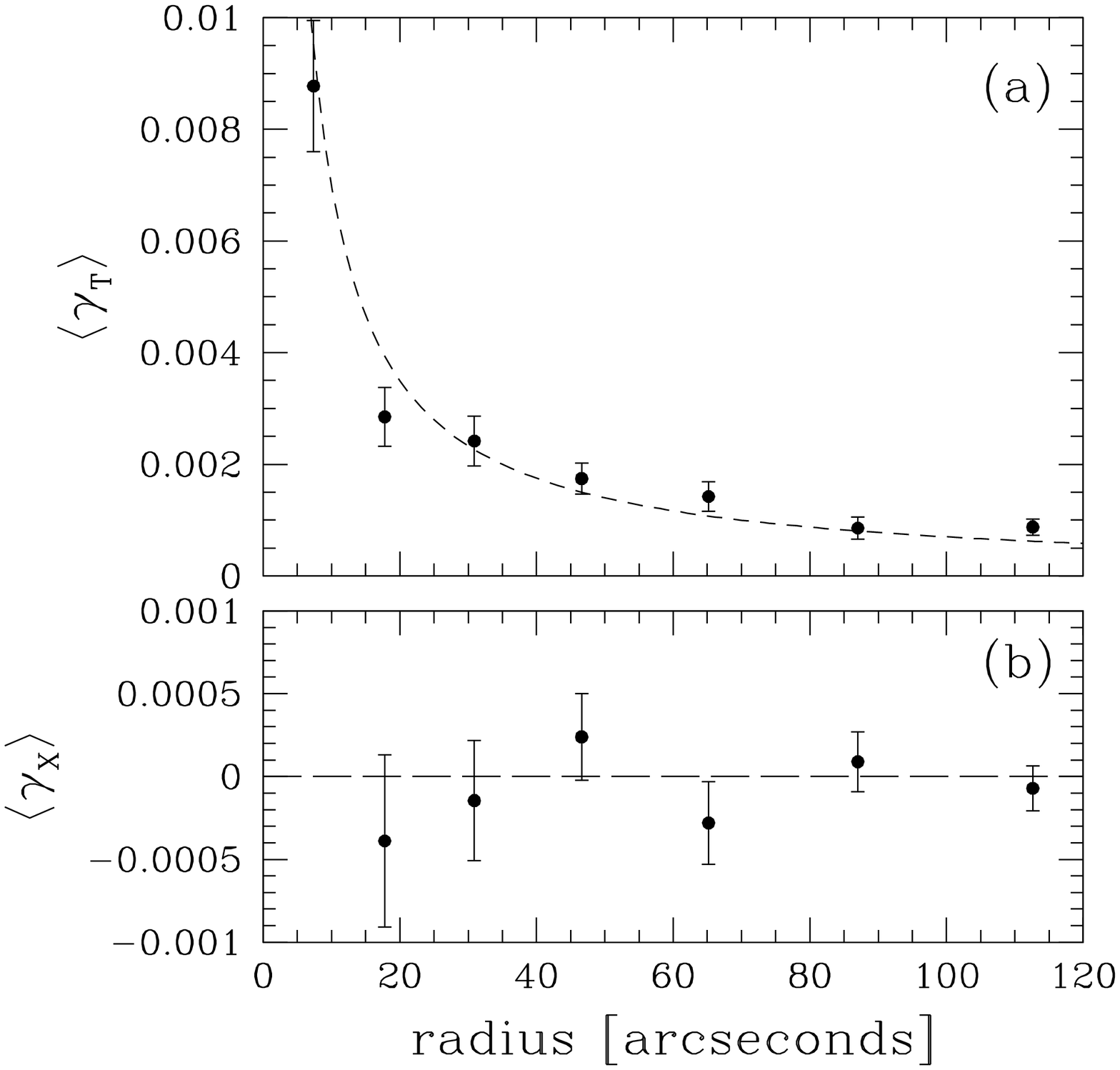}}
\figcaption{\footnotesize (a) Ensemble averaged tangential shear as a
function of radius out to 2 arcminutes from the lens. The solid line
corresponds to the best fit SIS model to the data at radii smaller
than 2 arcminutes. (b) The signal when the phase of the shear is increased
by $\pi/4$. Note the different vertical scale between panels a and b.
\label{gtprofzoom}}
\end{center}}

The signal on small angular scales is dominated by a single lens
galaxy, and can be used to obtain an estimate of the mass weighted
velocity dispersion of the sample of lenses (although such a mass
estimate can still be slightly biased because of the clustering of the
lenses). Figure~\ref{gtprofzoom}a shows the ensemble averaged
tangential shear on small scales. The measurements presented in
Figure~\ref{gtprofzoom}b show no evidence for residual systematics.

We fit a singular isothermal sphere (SIS) model to the tangential
shear at radii smaller than 2 arcminutes (which corresponds to $\sim
350 h^{-1}$ kpc at the mean redshift of the lenses).  The best fit
model is indicated by the dashed line in Figure~\ref{gtprofzoom}a.
For the Einstein radius $r_E$ we obtain a value of $\langle r_{\rm
E}\rangle=0\farcs{140} \pm0\farcs{009}$. If we extend the fit to much
larger radii, the inferred value for the Einstein radius increases
systematically. With our adopted redshift distributions for the lenses
and the sources, the value of $r_{\rm E}$ corresponds to a value of
$\langle\sigma^2\rangle^{1/2}=128\pm4$ km/s. The corresponding
circular velocity can be obtained using $V_c=\sqrt{2}\sigma$ (for a
spherical halo). The derived value of $\langle\sigma^2\rangle^{1/2}$
depends on the selection of the sample of lens galaxies, which hampers
a direct comparison with other studies.

It is more useful to compute the velocity dispersion of a galaxy with
some fiducial luminosity, for which we take \lumstar. To do so, we
have to adopt a scaling relation between the velocity dispersion and
the luminosity. We assume $\sigma\propto L_B^{0.3}$ (see
\S4). Furthermore, the luminosity might evolve with redshift, and we
will consider two cases. Under the assumption that the luminosity does
not evolve with redshift, we obtain a velocity dispersion
$\sigma=140\pm 4 \pm 3$ km/s for a galaxy with \lumstar. Hoekstra et
al. (2003) measured $\sigma=115^{+15}_{-17}$ km/s for a galaxy with a
luminosity of \lumhoek using a sample of galaxies with redshifts from
CNOC2. Scaling our measurements to this luminosity implies a velocity
dispersion of $118\pm4\pm2$ km/s, in excellent agreement with Hoekstra et
al. (2003). If we assume that the luminosity scales with redshift as
$L_B\propto(1+z)$, we find a velocity dispersion of $\sigma=150\pm 4\pm3$
km/s for a galaxy with \lumstarz.

\section{Properties of dark matter halos}

The galaxy-mass cross-correlation function is the convolution of the
galaxy distribution and the galaxy dark matter profiles. To examine
the ensemble average properties of the dark matter halos properly, we
need to deconvolve the galaxy-mass cross-correlation function (i.e., we
need to account for the clustering of the lenses).  This is done
naturally in a maximum likelihood analysis, where a model for the mass
distribution of individual galaxies is compared to the observations.

Fisher et al. (2000) and McKay et al. (2001) used a model with the
velocity dispersion and ``extent'' of the halo as free parameters, and
computed the resulting galaxy-mass correlation function for a range of
model parameters and compared the results to the observed galaxy-mass
correlation function. They obtained good constraints for the velocity
dispersion, but were not able to constrain the extent of the dark
matter halos.

The galaxy-mass correlation function ignores additional information
contained in the data, whereas the maximum likelihood analysis allows
one to gauge how much signal might arise from nearby
galaxies. Hoekstra et al. (2003) compared their mass model to the
observed polarisation field, making use of both components of the
polarisation (also see Brainerd et al. 1996; Schneider \& Rix 1997;
Hudson et al. 1998). This greatly improved the constraints on the
extent of the halo, and we will use the same method here.

In our analysis we make an important assumption: all clustered matter
is associated with the lenses. If the matter in galaxy groups (or
clusters) is associated with the halos of the group members (i.e., the
halos are indistinguishable from the halos of isolated galaxies) our
results should give a fair estimate of the extent of galaxy
halos. However, if a significant fraction of the dark matter is
distributed in common halos, a simple interpretation of the results
becomes more difficult. 

Hoekstra et al. (2003) examined how known groups in their observed
fields affected the lensing results, and found that the masses and
sizes might be overestimated by at most $\sim 10\%$. Guzik \& Seljak
(2002) found similar results from their analysis of the galaxy-galaxy
lensing signal in the context of halo models. Their approach allows
one to separate the contribution from groups to the lensing signal. As
expected, Guzik \& Seljak (2002) found that the effect depends on
galaxy type: early type galaxies are found in high density regions,
and are affected more. Alternatively, a comparison with numerical
simulations which include a prescription for galaxy formation (e.g.,
Kauffmann et al. 1999a, 1999b; Guzik \& Seljak 2001) can be used to
quantify this effect.

Another complication is the fact that we cannot separate the lenses in
different morphology classes with the current data. Early and late
type galaxies of a given luminosity have different masses, etc. (e.g.,
Guzik \& Seljak 2002). Hence, it is important to keep in mind that the
results presented here are ensemble averages over all galaxy
types. This is where the SDSS can play an important role, although we
can significantly improve the RCS results with upcoming multi-color
data.

The variance in the polarisations is approximately constant with
apparent magnitude and we approximate the distribution by a Gaussian
distribution. With the latter assumption, the log-likelihood follows a
$\chi^2$ distribution with the number of degrees of freedom equal to
the number of free model parameters, and the determination of
confidence intervals is straightforward. The log-likelihood is given
by the sum over the two components of the polarisation $e_i$ of all
the source galaxies

\begin{equation}
\log{\cal L}=-\sum_{i,j} \left(\frac{e_{i,j} - P^\gamma_j
g_{i,j}^{\rm model}}{\sigma_{e_j}}\right)^2,
\end{equation}

\noindent where $g_{i,j}$ are the model distortions, $P^\gamma_j$ is
the shear polarisability, $e_{i,j}$ are the observed image
polarisations for the $j$th galaxy, and $\sigma_{e_j}$ is the
dispersion in the polarisation of the $j$th galaxy (which is the
combination of the intrinsic shape of the galaxy and the measurement
error caused by shot noise in the images). In order to minimize the
contribution of the baryonic (stellar) component of the galaxies we
compare our model to the observations at radii larger than 10
arcseconds (which corresponds to $\sim 30 h^{-1}$ kpc at the redshift
of the lenses), where the dark matter halo should dominate the lensing
signal.

In our maximum likelihood analysis we ignore the contribution from
lenses outside the field of view (e.g., Hudson et al. 1998). For small
fields of view this tends to slightly lower the resulting halo masses
and sizes. The area covered by our observations is much larger than
the HDF North studied by Hudson et al. (1998), and the effect on our
estimates is negligible.

To infer the best estimates for the model parameters, one formally has
to perform a maximum likelihood analysis in which the redshift of each
individual galaxy is a free parameter, which has to be chosen such
that it maximizes the likelihood. This approach is computationally not
feasible, and instead we create mock redshift catalogs, using the
observed redshift distributions from the CNOC2 survey (Hoekstra et al.
2003), which allows us to find estimates for the model parameters.

For each lens galaxy in the RCS catalog, with a given apparent $R_C$
magnitude, we randomly draw a galaxy from the CNOC2 survey (with a
similar $R_C$, but allowing for a small range in magnitude). The lens
galaxy is then assigned the redshift and restframe $B$-band luminosity
of that CNOC2 galaxy. We take the incompleteness of the survey into
account when the redshifts are drawn from the survey. The mock
catalogs are then analysed as if the redshifts of the lenses were
known. Although this procedure does not provide the formal maximum
likelihood parameter estimation, it does yield an unbiased estimate of
the parameters. The procedure is repeated 10 times and enables us to
properly account for the uncertainty introduced by the lack of precise
redshifts for the lenses.

Future multi-color catalogs will improve the results presented here
significantly. In particular, in the analysis presented here we have
to adopt scaling relations between the velocity dispersion (or mass)
and the luminosity of the lens and the extent of the halo and the
luminosity. The use of photometric redshifts for the lenses will allow
us to actually contrain these relations.

In \S4.1 we consider the Truncated Isothermal Sphere (TIS; e.g.,
Brainerd et al. 1996; Schneider \& Rix 1997; Hoekstra et al. 2003),
which has been used previously in galaxy-galaxy lensing analyses. In
\S4.2 we compare the data to the popular NFW profile (Navarro et
al. 1995, 1996, 1997).

\subsection{Truncated Isothermal Sphere Model}

A simple model is the truncated isothermal sphere proposed by Brainerd
et al. (1996). Its density profile is given by

\begin{equation}
\rho(r)=\frac{\sigma^2 s^2}{2\pi G r^2(r^2+s^2)},
\end{equation}

\noindent where $\sigma$ is the line-of-sight velocity dispersion, and
$s$ is a truncation scale, i.e. the radius where the profile steepens.
On small scales $(r\ll s)$ the model behaves as an Singular Isothermal
Sphere (SIS) model, whereas for $r\gg s$ the density decreases
$\propto r^{-4}$. The mass contained within a sphere of radius 
$r$ is given by 

\begin{equation}
M(r)=\frac{2\sigma^2 s}{G} \arctan(r/s),
\end{equation}

\noindent which results in a finite total mass of 

\begin{equation}
M_{\rm tot}=\frac{\pi\sigma^2}{G}s=7.3\times 10^{12}
\left(\frac{\sigma}{{\rm 100~km/s}}\right)^2
\left(\frac{s}{{\rm 1~Mpc}}\right).
\end{equation}

\noindent The projected surface density for this model is given by

\begin{equation}
\Sigma(r)=\frac{\sigma^2}{2Gr}\left(1-\frac{r}{\sqrt{r^2+s^2}}\right).
\end{equation}

\noindent The corresponding expressions for the shear can be found in
Brainerd et al. (1996) and Schneider \& Rix (1997).

We use this model to compute the model shear field and compare it to
the data. The lenses, however, span a range in masses and we need to
account for that using scaling relations, which allow us to relate the
halo properties of the lenses to those of a fiducial galaxy. For the
fiducial galaxy we take a luminosity of \lumstar.

Dynamical studies provide evidencence of a power law scaling relation
between the velocity dispersion and the luminosity (e.g., Tully-Fisher
relation for spiral galaxies and Faber-Jackson relation for early type
galaxies). We assume $\sigma\propto L_{\rm B}^{0.3}$, which is based
on the observed slope of the $B$-band Tully-Fisher relation (e.g.,
Verheijen 2001). Little is known, however, about the relation of the
extent of dark matter halos with other (observable) parameters. Using
SDSS data, Guzik \& Seljak (2002) find that $M\propto
L_{g'}^{1.2\pm0.2}$. Motivated by this result, we adopt $s\propto
L_B^{0.6}$, which gives a total mass $M\propto L_B^{1.2}$. We note,
however, that we probe lower luminosities than Guzik \& Seljak (2002)
(also see McKay et al. 2001) and as a result, the adopted scaling
relation might differ from the actual ones.

\vbox{
\begin{center}
\leavevmode 
\hbox{% 
\epsfxsize=8.5cm 
\epsffile{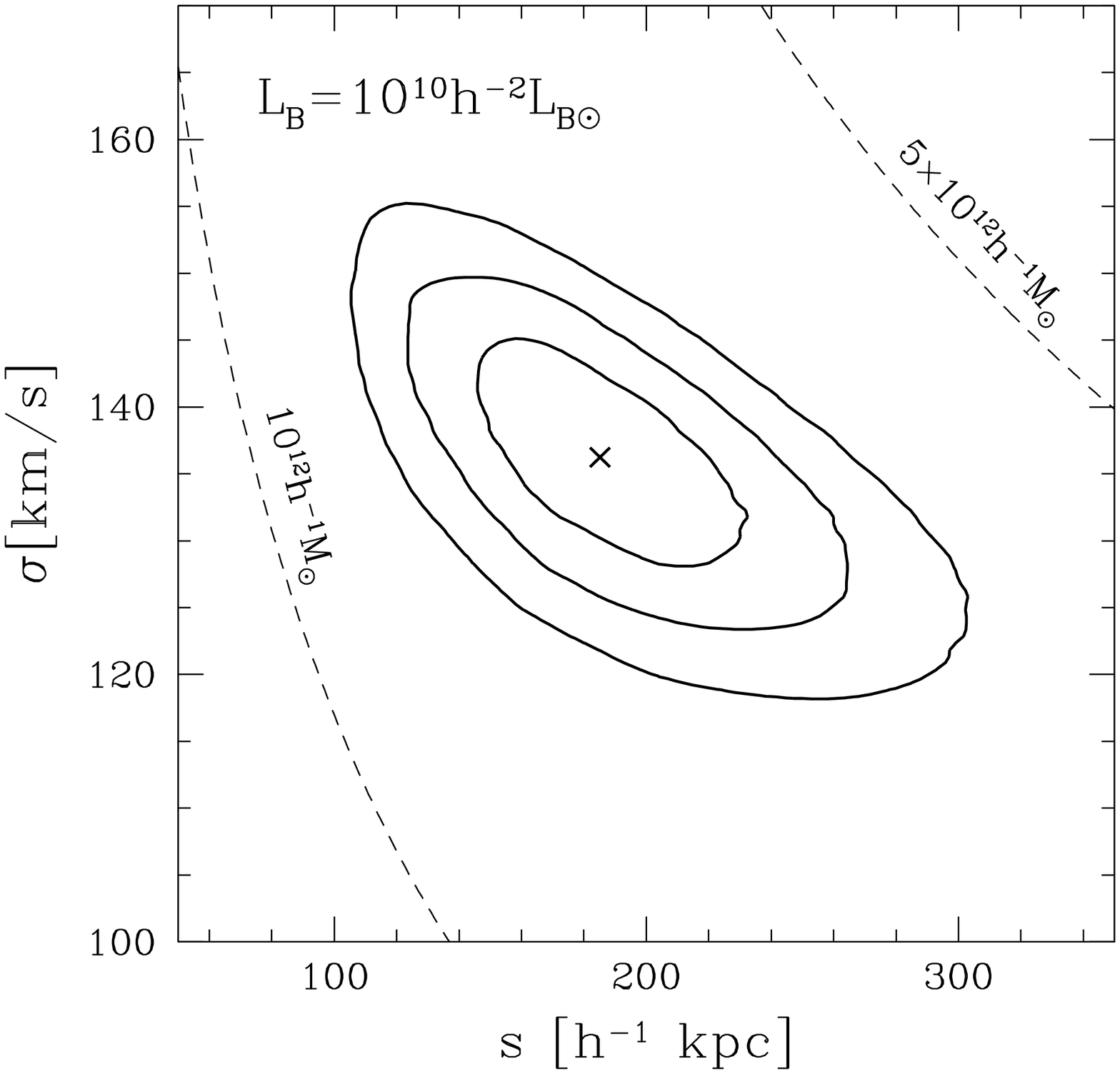}}
\figcaption{\footnotesize Joint constraints on the velocity dispersion
$\sigma$ and truncation parameter $s$ for a fiducial galaxy with
$L_{\rm B}=10^{10}h^{-2}L_{{\rm B}\odot}$. The contours indicate the
68.3\%, 95.4\%, and the 99.7\% confidence on two parameters
jointly. The cross indicates the best fit value. The dashed lines
indicate models with masses $M_{\rm tot}=10\times 10^{12} h^{-1} M_\odot$
and $M_{200}=5 \times 10^{12} h^{-1} M_\odot$.
\label{size}}
\end{center}}

Figure~\ref{size} shows the joint constraints on the velocity
dispersion $\sigma$ and truncation parameter $s$ for a fiducial galaxy
with \lumstar, under the assumption that the luminosity does not
evolve with redshift. Conveniently, for this particular luminosity,
the inferred values of the velocity dispersion and truncation
parameter depend only marginally on the adopted scaling relations.

For the velocity dispersion we obtain a value of $\sigma=136 \pm 5 \pm
3$ km/s (68\% confidence, marginalizing over all other model
paremeters).  Hoekstra et al. (2003) find a velocity dispersion of
$\sigma=110\pm 12$ for a galaxy with a luminosity of \lumhoek. For a
galaxy with that luminosity, the adopted scaling relations imply a
velocity dispersion of $114\pm 4\pm 2$ km/s, in excellent agreement
with Hoekstra et al. (2003).  McKay et al. (2001) who found a best fit
value of $\sigma=113^{+17}_{-13}$ km/s (95\% confidence) for a galaxy
with $L_{g'}\sim 9\times 10^9 h^{-2} L_{g'\odot}$. Our results
correspond to a velocity dispersion of $\sigma=127\pm5\pm3$ km/s for a
galaxy of that luminosity, in good agreement with the SDSS result.

We derive tight constraints on the truncation parameter, i.e. the
extent of dark matter halos. We find a value of $s=185^{+30}_{-28}
h^{-1}$ kpc (68\% confidence), and a total mass $M_{\rm
tot}=(2.5\pm0.3\pm0.1)\times 10^{12} h^{-1} M_\odot$. The results
presented in Hoekstra et al. (2003) imply a value of
$s=290^{+139}_{-82} h^{-1}$ kpc (68\% confidence) for their fiducial
galaxy. For a galaxy with a luminosity of \lumhoek, we obtain
$s=131^{+21}_{-20} h^{-1}$ kpc, marginally consistent with Hoekstra et
al. (2003).

If we assume that $L_B\propto(1+z)$ we obtain a velocity dispersion of
$\sigma=146\pm5\pm3$ km/s and a truncation size of $s=213^{+35}_{-32}
h^{-1}$ kpc for a galaxy with \lumstarz.

\subsection{NFW model}

Numerical simulations of collisionless cold dark matter (CDM)
reproduce the observed structure in the universe remarkably
well. Furthermore these simulations suggest that CDM gives rise to a
specific density profile, which fits the radial mass distribution for
halos with a wide range in mass (e.g., Dubinski \& Carlberg 1991;
Navarro et al. 1995, 1996, 1997). The NFW density profile is
characterized by 2 parameters, a density contrast $\delta_c$ and a
scale $r_s$

\begin{equation}
\rho(r)=\frac{\delta_c\rho_c}{(r/r_s)(1+r/r_s)^2},
\end{equation}

\noindent where $\rho_c$ is the critical surface density at the
redshift of the halo. The ``virial'' radius $r_{200}$ is defined as
the radius where the mass density of the halo is equal to $200\rho_c$, 
and the corresponding mass $M_{200}$ inside this radius is given by

\begin{equation}
M_{200}=\frac{800\pi}{3}\rho_c r_{200}^3,
\end{equation}

\noindent with a corresponding rotation velocity $V_{200}$ of

\begin{equation}
V_{200}=V(r_{200})=\frac{G M_{200}}{r_{200}}.
\end{equation}

\noindent The concentration parameter is defined as $c=r_{200}/r_s$,
which yields an expression for the overdensity of the halo $\delta_c$
in terms of $c$

\begin{equation}
\delta_c=\frac{200}{3}\frac{c^3}{\ln(1+c)-c/(1+c)}.
\end{equation}

It is important to note that the density profile on small scales
remains controversial as other groups find different results (e.g.,
Moore et al. 1999; Ghigna et al. 2000). In addition, realistic
simulations should include the effect of baryons, which complicate
matters even further (see e.g., Mo, Mao \& White 1998; Kochanek \&
White 2001). Unfortunately the current RCS data do not allow us to
constrain the slope of the inner mass profile. However, future, deep
lensing surveys, such as the CFHT Legacy Survey, will be well suited
for such a study.

On the other hand, there is good agreement for the density profile on
large scales, and a comparison of the profiles of real objects with
the predictions provides an important test of the assumption that
structures form through dissipationless collapse.  The predicted
profiles agree well with the observed mass distribution in clusters of
galaxies (e.g., Hoekstra et al. 2002d), but the situation is less
clear for galaxy mass halos. Rotation curves can provide some
constraints, but typical values for $r_s$ for galaxy mass halos are
$10-20 h^{-1}$ kpc, comparable to the outermost point for which
rotation curves have been measured.

Galaxies that are thought to be dark matter dominated, such as low
surface brightness galaxies, potentially might be more suitable to
test the CDM predictions.  Studies of the rotation curves of low
surface brightness galaxies suggest that, at least for a fraction of
them, the observed rotation curves rise more slowly than the CDM
predictions (e.g, de Blok, McGaugh \& Rubin 2001; McGaugh, Barker \&
de Blok 2002).  It is not clear, however, whether such studies provide
a good test of CDM, because low surface brightness galaxies are
peculiar (Zwaan \& Briggs 2000), and their formation is not well
understood. Hence, it is not obvious that their halos should be
described by an NFW profile. Recently Ricotti (2003) has suggested
that the inner slope might depend on halo mass, with low mass systems
having shallow cores, whereas massive galaxies are well described by
the NFW profile.

In this section we compare the NFW profile to the observations, with
$\delta_c$ and $r_s$ as free parameters. The equations describing the
shear for the NFW profile have been derived by Bartelmann (1996) and
Wright \& Brainerd (2000). 

As before we have to adopt scaling relations. We assume that the
maximum rotation velocity scales $\propto L_B^{0.3}$ (the B-band
Tully-Fisher relation; Verheijen 2001). If we also assume that
$M_{200}\propto L_B^{1.2}$, as motivated by the findings of Guzik \&
Seljak (2002), we obtain that $r_s\propto L_B^{0.75}$ and $\delta_c
\propto L_B^{-0.85}$.

Figure~\ref{size_nfw} shows the joint constraints on $V_{200}$ and
$r_s$ for a galaxy with a luminosity of \lumstar. In addition, the
right axis indicates the corresponding values for $M_{200}$. We derive
a best fit value of $V_{200}=162\pm5\pm3$ km/s, or
$M_{200}=(8.4\pm0.7\pm0.4)\times 10^{11} h^{-1} M_\odot$ (68\% confidence),
and a corresponding value of $r_{200}=139^{+3}_{-5} h^{-1}$ kpc.  It
is useful to compare this result with the mass from the TIS model.

The TIS model yields $M_{\rm TIS}(r_{200})=(1.0\pm0.1)\times 10^{12}
h^{-1} M_\odot$, which is slightly larger than the NFW value. As shown
by Wright \& Brainerd (2000), isothermal models give higher masses
compared to NFW models. Hence, the results derived from both models
are consistent.

From their galaxy-galaxy lensing analysis of the SDSS, Guzik \& Seljak
(2002) find $M_{200}=(9.3\pm1.6)\times 10^{11} h^{-1} M_\odot$ for a
galaxy of $L_{g'}\sim 1.1\times 10^{10} h^{-2} L_{g'\odot}$, in good
agreement with our results.

For the scale $r_s$ we find $r_s=16.2^{+3.6}_{-2.9} h^{-1}$ kpc (68\%
confidence), and the best fit density contrast is
$\delta_c=2.4^{+1.4}_{-0.8}\times 10^4$ (68\% confidence; confidence
interval from Monte Carlo simulation). The TIS model provides a
slightly better fit to the data, but the difference is not
significant, and consequently the data are not sufficient to
distinguish between the NFW and TIS model.

As before, we also calculated the results under the assumption that
the luminosity evolves $\propto (1+z)$. In this case we find a value
of $r_s=17.2^{+3.8}_{-3.1} h^{-1}$ kpc and $V_{200}=176\pm 5\pm 4$
km/s for a \lumstarz galaxy.

In our maximum likelihood analysis we considered $r_s$ and $V_{200}$
(or equivalently the mass $M_{200}$) as free parameters. Numerical
simulations of CDM, however, show that the parameters in the NFW model
are correlated, albeit with some scatter. As a result, the NFW model
can be considered as a one-parameter model: given the cosmology,
redshift, and one of the NFW parameters, the values for all other
parameters can be computed using the routine {\tt CHARDEN} made
available by Julio Navarro\footnote{The routine {\tt CHARDEN} can be
obtained from {\tt http://pinot.phys.uvic.ca/${\tilde{\
}\!}$jfn/charden}}.  Hence, the simulations make a definite prediction
for the value of $V_{200}$ as a function of $r_s$. The dotted
line in Figure~\ref{size_nfw} indicates this prediction. If the
simulations provide a good description of dark matter halos, the
dotted line should intersect our confidence region, which it does.

This result provides important support for the CDM paradigm, as it
predicts the correct ``size'' of dark matter halos. It is important to
note that this analysis is a direct test of CDM (albeit not
conclusive), because the weak lensing results are inferred from the
gravitational potential at large distances from the galaxy center,
where dark matter dominates.  Most other attempts to test CDM are
confined to the inner regions, where baryons are, or might be,
important.

\vbox{
\begin{center}
\leavevmode 
\hbox{% 
\epsfxsize=8.5cm 
\epsffile{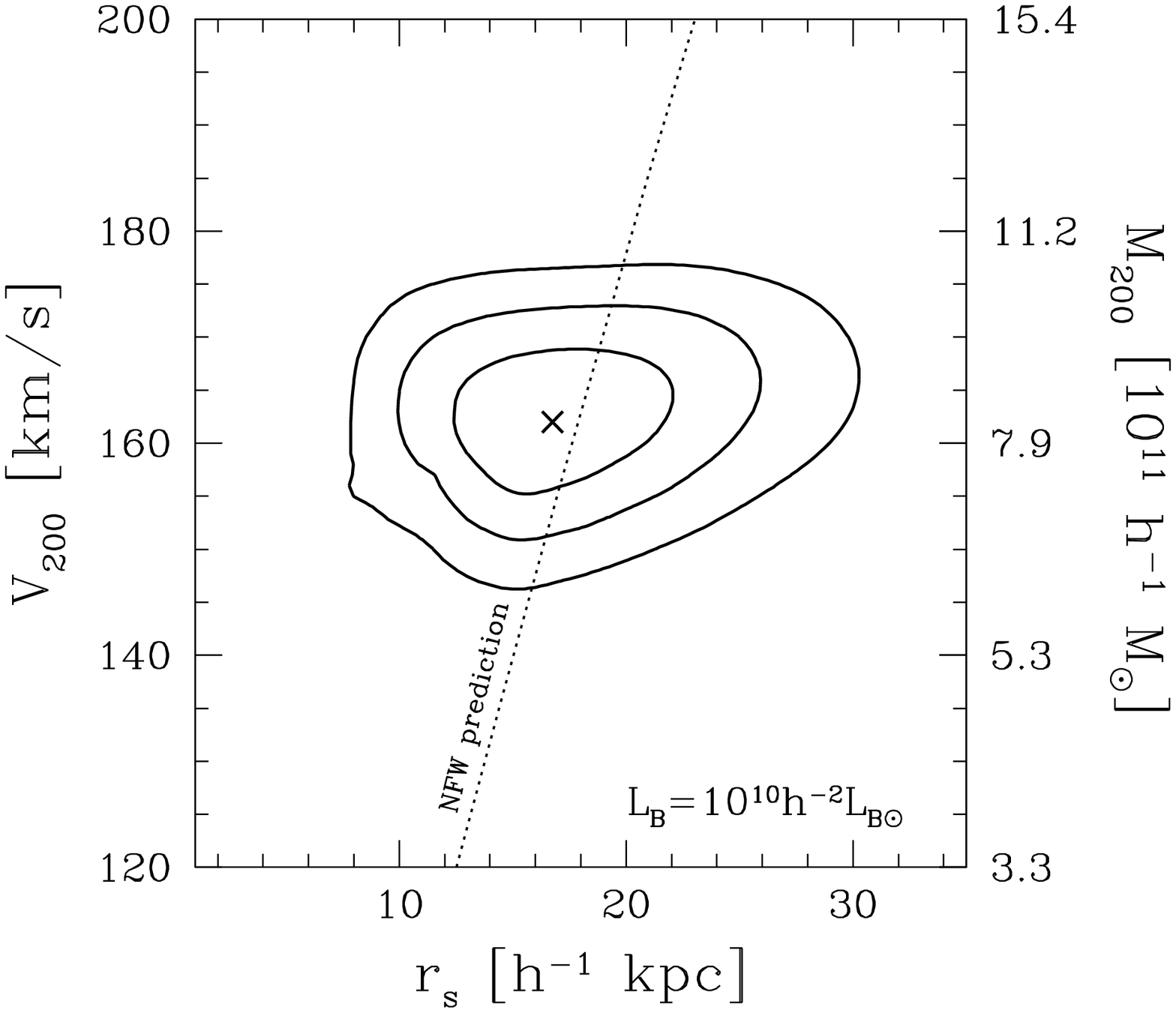}}
\figcaption{\footnotesize Joint constraints on $\delta_c r_s^2$ and
scale radius $r_s$ for a fiducial galaxy with $L_{\rm
B}=10^{10}h^{-2}L_{{\rm B}\odot}$, with an NFW profile. The contours
indicate the 68.3\%, 95.4\%, and the 99.7\% confidence on two
parameters jointly. The cross indicates the best fit value.
The dashed lines indicate models with masses $M_{200}=5\times 10^{11}
h^{-1} M_\odot$ and $M_{200}=10^{12} h^{-1} M_\odot$. The dotted
line indicates the predictions from the numerical simulations, which
are in excellent agreement with our results.
\label{size_nfw}}
\end{center}}

\section{Shapes of halos}

The average shape of dark matter halos can provide important
information about the nature of dark matter. Numerical simulations of
cold dark matter yield triaxial halos, with a typical ellipticity of
$\sim 0.3$ (e.g., Dubinski \& Carlberg 1991).  Hence, in the context
of collisionless cold dark matter, the theoretical evidence for
flattened halos is quite strong. If the dark matter is interacting, it
tends to produce halos that are more spherical (compared to cold dark
matter). This difference is more pronounced in the central parts of
the halo, where the density is high.  On the large scales probed by
weak lensing, the different types of dark matter (for reasonable
interaction cross-sections) produces halos with similar shapes.

Nevertheless, a measurement of the average shape of dark matter halos
is important, because the observational evidence is still limited.
Dynamical measurements are limited by the lack of visible tracers, and
therefore only probe the vertical potential on scales $\le$ 15 kpc.
Although the spread in inferred values for the axis ratio $c/a$ (where
$c/a$ is the ratio of the shortest to longest principle axis of the
halo) is large, the results suggest an average value of
$c/a=0.5\pm0.2$ (Sackett 1999).

Weak gravitational lensing is potentially the most powerful way to
derive constraints on the shapes of dark matter halos.  The amount of
data required for such a measurement, however, is large (e.g.,
Brainerd \& Wright 2000; Natarajan \& Refregier 2000): the
galaxy-galaxy lensing signal is small, and now one needs to measure an
even smaller azimuthal variation. We also have to assume that the
galaxy and its halo are aligned. An imperfect alignment between light
and halo will reduce the amplitude of the azimuthal variation
detectable in the weak lensing analysis.  Hence, weak lensing formally 
provides a lower limit to the average halo ellipticity.

Brainerd \& Wright (2000) and Natarajan \& Refregier (2000) proposed
to study the azimuthal variation in the tangential shear around the
lenses. On very small scales, the lensing signal is dominated by the
lens, but on larger scales, the clustering of the lenses will lower
the signal one tries to measure (the two point function is
axisymmetric). We therefore use the maximum likelihood approach used
in the previous section.

To maximize the signal-to-noise ratio of the measurement one has to
assign proper weights to the lenses: edge-on galaxies have maximal
weight, whereas the lensing signal around face-on galaxies contains no
information about the shape of the halo. We adopt a simple approach,
and assume that the (projected) ellipticity of the dark matter halo is
proportional to the shape of the galaxy: $e_{\rm halo}=f e_{\rm
lens}$.

The measurement of the azimuthally averaged tangential shear around
galaxies is robust against residual systematics (e.g., imperfect
correction for PSF anisotropy): contributions from a constant or
gradiant residual shear cancel. This is no longer the case for the 
quadrupole signal, and imperfect correction for the PSF anisotropy
can mimick the signal from a flattened halo.

If the lens galaxy is oriented randomly with respect to the residual
shear, the average over many lenses will cancel the contribution from
systematics. In real data, however, the uncorrected shapes of the
lenses are aligned with the systematic signal. Hence, an imperfect
correction can give rise to a small quadrupole signal, although we
note that the lenses used in our analysis are large compared to the
PSF. We estimate the amplitude of this effect in Appendix~A, and show
that it is negligible for the measurements presented here. We also
examined the robustness of our results by splitting the data into two
samples and comparing the results.

\vbox{
\begin{center}
\leavevmode 
\hbox{% 
\epsfxsize=8.5cm 
\epsffile{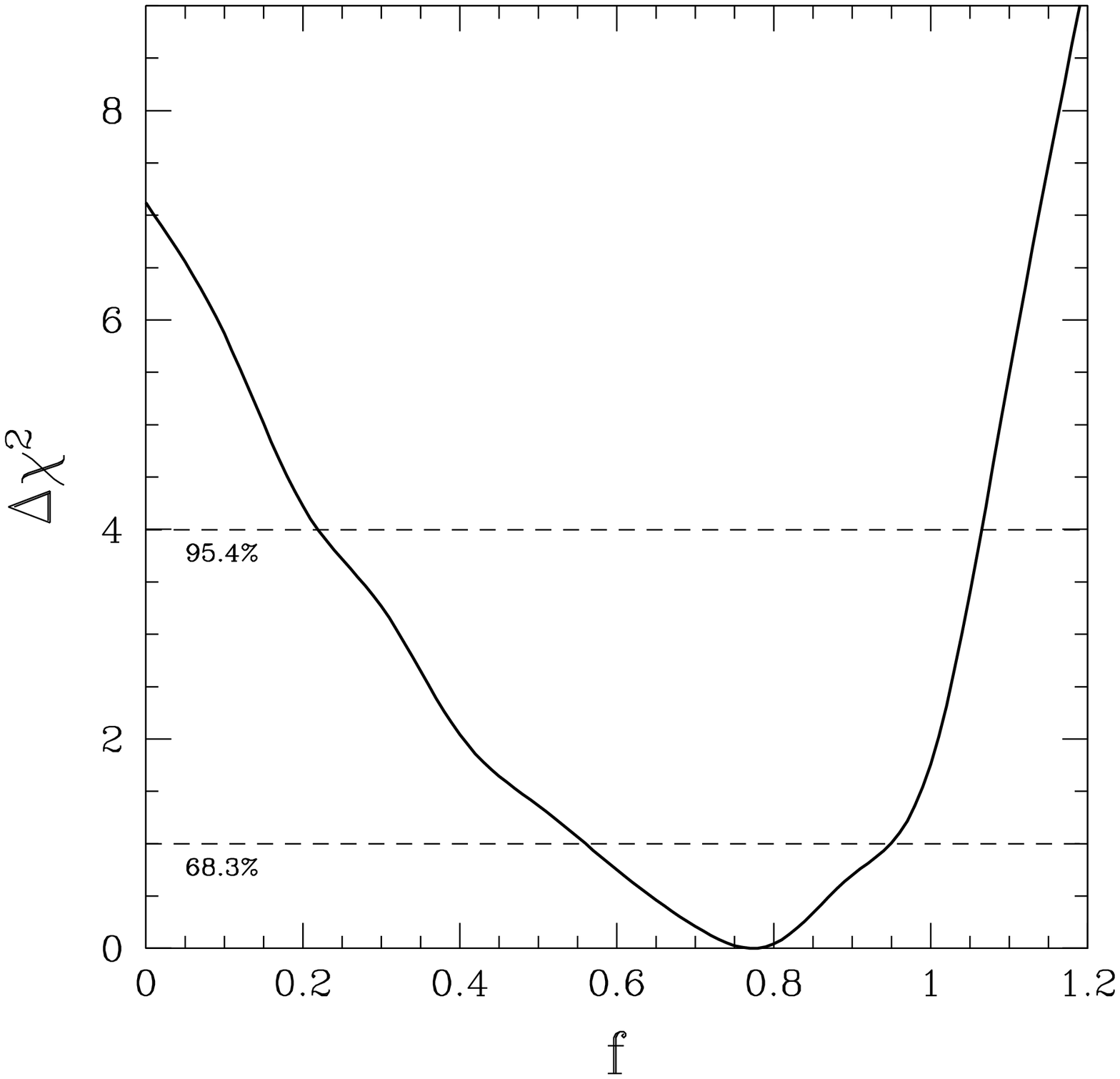}}
\figcaption{\footnotesize $\Delta \chi^2$ as a function of $f$. We have
assumed that the ellipticity of the halos is related to the observed
ellipticity of the lens as $e_{\rm halo}=f e_{\rm lens}$.  We have
indicated the 68.3\% and 95.4\% confidence intervals. We find a best
fit value of $f=0.77^{+0.18}_{-0.21}$ (68\% confidence). Round halos
$(f=0)$ are excluded with 99.5\% confidence. 
\label{flat_halo}}
\end{center}}

We use an elliptical TIS model to compute the model shear field, and
compare this to the data. Figure~\ref{flat_halo} shows the resulting
$\Delta\chi^2$ as a function of $f$.  We find a best fit value of
$f=0.77^{+0.18}_{-0.21}$ (68\% confidence). This suggests that, on
average, the dark matter distribution is rounder than the light
distribution. As discussed above, our analysis formally provides only
a lower limit on the halo ellipticity, and the true ellipticity might
be higher if some of the halos are misaligned with the light.
Nevertheless, the fact that we detect a significant flattening implies
that the halos are well aligned with the light distribution. Also note
that the lensing signal is caused by a range of different galaxy
types, for which our simple relation between the halo ellipticity and
light distribution might not be valid.

Consequently the interpretation of the results is difficult, although
a simple interpretation actually yields sensible results. For
instance, the average ellipticity of the lens galaxies is $\langle
e_{\rm lens}\rangle=0.414$. Hence, the measured value of $f$ implies
an average projected halo ellipticity of $\langle e_{\rm
halo}\rangle=0.33^{+0.07}_{-0.09}$ (68\% confidence), which
corresponds to an projected axis ratio of $c/a=0.67^{+0.09}_{-0.07}$
(68\% confidence; where we have used $c/a=1-e$). Although the weak
lensing yields a projected axis ratio, the result is in fair agreement
with the results from numerical simulations.

A robust outcome of our analysis is that spherical halos ($f=0$) are
excluded with 99.5\% confidence. As we demonstrate below, this poses
serious problems for alternative theories of gravity, which attempt to
explain the observations without dark matter.

\subsection{Implications for alternative theories of gravity}

In this section we examine the implications of our measurement of the
anisotropy in the lensing signal around galaxies for theories of
gravity without dark matter.  We focus on one particular approach:
Modified Newtonian Dynamics (MOND; Milgrom 1983; Sanders 1986; Sanders
\& McGaugh 2002), which has been shown to describe rotation curves
rather well (e.g., Begeman et al. 1991; Sanders \& Verheijen 1998).

In principle weak lensing can be used as a powerful test of MOND, but
unfortunately no relativistic description of MOND has been found.
Consequently one cannot compute the lensing signal in this theory.
However, even in the absence of an appropriate description of lensing,
we can use the observed anisotropy in the lensing signal around
galaxies to test MOND.

In any reasonable alternative theory of gravity, the anisotropy in the
lensing signal of an {\it isolated} galaxy is caused by the
distribution of light and gas in that galaxy. In order to explain the
flat rotation curves, these alternative theories typically need an
effective force law $\propto r^{-1}$. Hence, on small scales one
expects an anisotropic signal, but at large radii (where there are no
stars and gas) we assume that the anisotropy in the lensing signal
decreases $\propto r^{-2}$. As a result, these theories predict an
almost isotropic weak lensing signal on the scales probed by our
analysis, which is not observed.

Galaxies, however, are not isolated and the external field effect
(Milgrom 1986) might complicate the interpretation of our
measurements.  In MOND, if a galaxy is embedded in an external field,
this field dominates the dynamics if its acceleration is larger than
the acceleration of the galaxy. As a result, the effective
gravitational field is non-spherical, even if the potential of the
galaxy is isotropic (as is the case for an isolated galaxy).

This effect is important if galaxies would be aligned with this
external field. We know, however, that the intrinsic alignments of
galaxies are small (e.g., Lee \& Pen 2001, 2002) and for the
measurements presented here, it is safe to assume that the lenses have
random orientation with respect to any external field. Consequently,
the observed anisotropy in the lensing signal cannot be caused by the
external field effect.

Hence, our findings provide strong support for the existence of dark
matter, because alternative theories of gravity predict an almost
isotropic lensing signal. Better constraints can be derived from
future weak lensing surveys, which will allow us to study the
anisotropy a function of projected distance from the galaxy.

\section{Conclusions}

We have analysed the weak lensing signal caused by a sample of lenses
with $19.5<R_C<21$ using 45.5 deg$^2$ of $R_C$ band imaging data from
the Red-Sequence Cluster Survey (RCS). We have studied the average
mass profile around the lenses using a maximum likelihood analysis.
To this end, we considered two models for the halo mass profile: a
truncated isothermal sphere and an NFW profile. We have assumed
(observationally motivated) scaling relations between the luminosity
of the lens and the velocity dispersion and the extent of the halo.

The TIS model yields a best fit velocity dispersion of
$\sigma=136\pm5\pm 4$ km/s and a truncation radius $s=185^{+30}_{-28}
h^{-1}$ kpc for a galaxy with a fiducial luminosity of
\lumstar. Alternatively, the best fit NFW model yields a mass
$M_{200}=(8.4\pm0.7\pm 4)\times 10^{11} h^{-1} M_\odot$ and a scale
radius $r_s=16.2^{+3.6}_{-2.9} h^{-1}$ kpc. This value for the scale
radius is in excellent agreement with predictions from numerical
simulations for a halo of this mass. 

We also present the first detection of the flattening of galaxy dark
matter halos from weak lensing. We use a simple model in which the
ellipticity of the halo is $f$ times the observed ellipticity of the
lens. We find a best fit value of $f=0.77^{+0.18}_{-0.21}$ (68\%
confidence), suggesting that the dark matter halos are somewhat
rounder than the light distribution. The fact that we detect a
significant flattening implies that the halos are aligned with the
light distribution. Given the average ellipticity of the lenses, this
implies a halo ellipticity of $\langle e_{\rm
halo}\rangle=0.33^{+0.07}_{-0.09}$ (68\% confidence), in fair
agreement with results from numerical simulations of CDM. This result
provides strong support for the existence of dark matter, as an
isotropic lensing signal is excluded with 99.5\% confidence.

\acknowledgments

We are grateful to the anonymous referee whose comments have
significantly improved the quality of this paper. The RCS project is
partially supported by grants from the Natural Science and Engineering
Science Council of Canada and the University of Toronto to HKCY.

\appendix
\section{Contribution of systematics to the shape measurement of dark halos}

In this appendix we examine how residual systematics affect the
measurement of the flattening of dark matter halos. A schematic
overview of the situation is presented in Figure~\ref{res_fig}.  
The thin lines indicate the direction of residual systematics.  The
residual shear has an amplitude $\hat\gamma$ and a position angle
$\phi$ with respect to the major axis of the lens. The tangential
shear $\gamma_T^{\rm obs}$ observed at a position $(r,\theta)$ is the
sum of the lensing signal $\gamma_T^{\rm lens}$ and the contribution
from systematics $\hat\gamma_T$. The latter is given by

\begin{equation}
\hat\gamma_T=-\hat\gamma[\cos(2\phi)\cos(2\theta)+\sin(2\phi)\sin(2\theta)]=
-\hat\gamma\cos(2(\theta-\phi)).
\end{equation}

Hence, the azimuthally averaged tangential shear is not affected by
systematics as $\int d\theta \hat\gamma_T(\theta)=0$, and the
weak lensing mass estimate is very robust. For a flattened halo, the
lensing signal $\gamma_T^{\rm lens}$ is given by

\begin{equation}
\gamma_T^{\rm lens}(r,\theta)=[1+\gamma_f\cos(2\theta)]\cdot\langle\gamma_T\rangle(r),
\end{equation}

\noindent where $\langle\gamma_T\rangle$ is the azimuthally 
averaged tangential shear, and $\gamma_f$ is a measure of the
flattening of the halo. For positive values of $\gamma_f$, the
halo is aligned with the lens.

We consider the worst case scenario, and demonstrate that even in this
situation the results are robust. One way to estimate the flattening
of the halo is to measure the shears $\gamma_+$ (at $\theta=0$ and
$\pi$), and $\gamma_-$ (at $\theta=\pi/2$ and $3\pi/2$). The observed
ratio $f=\gamma_-/\gamma_+$ is

\begin{equation}
f_{\rm obs}=\frac{\gamma_- +\hat\gamma \cos(2\phi)}{\gamma_+ -
\hat\gamma \cos(2\phi)}.
\end{equation}

If $\phi$ is uncorrelated with the lens, the observed ratio, averaged
over many lenses, is unaffected by systematics because
$\langle\cos(2\phi)\rangle=0$. However, in real data, the PSF
anisotropy affects both the lens and the source galaxies. Although the
lenses used in this paper are large compared to the PSF, any (small)
residual in the correction will introduce a correlation in the
position angle of the lens and the direction of the PSF anisotropy.

\vbox{
\begin{center}
\leavevmode 
\hbox{% 
\epsfxsize=8.5cm 
\epsffile{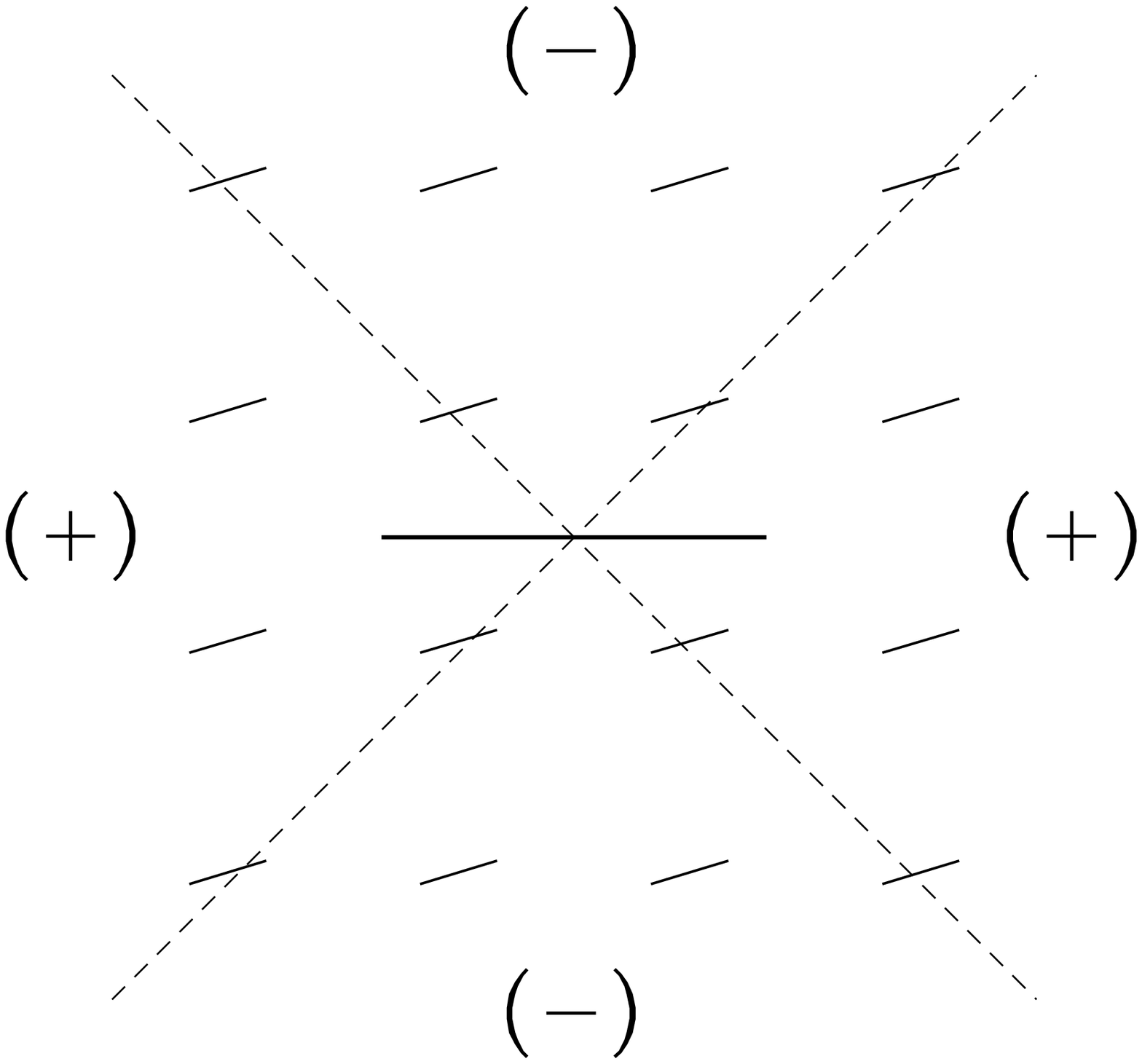}}
\figcaption{\footnotesize Schematic view of the lens galaxy and the
residual systematic shear. If the dark halo is flattened and aligned
with the lens (line in the center) then the tangential shear at a
given radius is larger in the quadrants indicated by (+), and lower in
(-). The small lines indicate the direction of the residual shear
(e.g., caused by the imperfect correction for PSF anisotropy). 
The residual shear has an amplitude $\hat\gamma$ and has a direction
with a position angle $\phi'$ with respect to the lens.
\label{res_fig}}
\end{center}}

The effect is maximal for $\phi=0$ (aligned with the major axis of the
lens) or $\phi=\pi/2$ (perpendicular to major axis of the lens). The
latter situation corresponds to an overcorrection of the PSF
anisotropy, whereas the former occurs when the correction for PSF
anisotropy is too small. For both (extreme) situations, only the
observed value of $\gamma_1$ is affected by systematics ($\gamma_2$ is
left unchanged). The observed value $\gamma_1^{\rm obs}$ for the lens
can be written as $\gamma_1^{\rm obs}=\gamma_1^{\rm true}+
\alpha\hat\gamma$, where $\alpha$ is a measure of the correlation
between the position angle of the lens and the direction of the
systematic shear of the background galaxies. In our case $\alpha\ll
1$, because the lenses are large compared compared to the PSF. For a
lens with with an observed $\gamma=\sqrt{\gamma_1^2+\gamma_2^2}$, we
find

\begin{equation}
\langle\cos(2\phi)\rangle=\frac{\alpha \hat\gamma}{2 \gamma}.
\end{equation}

The distribution of $\gamma_1$ and $\gamma_2$ can be approximated
by a Gaussian with a dispersion $\sigma$ (with a typical value of
$\sigma=0.2$). For the ensemble of lenses we then obtain

\begin{equation}
\langle\cos(2\phi)\rangle=\sqrt{\frac{\pi}{2}}\frac{\alpha\hat\gamma}{2\sigma}.
\end{equation}

This introduces a relatively large signal because face-on lenses
($\gamma\approx 0$) align easily with the PSF anisotropy. Such lenses,
however, contain no information about the shape of the halo. In our
analysis we assign a weight $\propto\gamma$ to each lens. Hence in our
case we are less sensitive to systematics as the correct estimate is
given by

\begin{equation}
\langle\cos(2\phi)\rangle=\frac{\alpha\hat\gamma}{2}.
\end{equation}

\noindent Hence, the observed ratio $\gamma_-/\gamma_+$ reduces to

\begin{equation}
f_{\rm obs}=\frac{\gamma_- + \hat\gamma^2\alpha/2}{\gamma_+ - \hat\gamma^2\alpha/2}
\end{equation}

We can now estimate how robust the measurement of the average halo
shape is. We take a conservative estimate of $\alpha=0.5$, and
$\hat\gamma=5\times 10^{-3}$ (see Hoekstra et al. 2002a for a
discussion of the residuals in our data). The average separation of
stars in our data is $\sim 1\farcm5$ and therefore the residual PSF
anisotropy is likely to change direction on scales larger than the
separation of the stars used to measure the anisotropy. The average
tangential shear at 2 arcminutes is $\sim 5\times 10^{-4}$. For a
spherical halo ($\gamma_-=\gamma_+= 5\times 10^{-4}$) we would observe
a ratio $f_{\rm obs}=1.025$. This corresponds to an ellipticity of
2.5\% (with the halo oriented perpendicular to the lens), which is
small ($\sim$ 10\% of the observed ellipticity).

In this very conservative estimate, we have assumed that for each
lens, the residual PSF anisotropy is aligned with the lens (i.e., the
PSF anisotropy was underestimated). There is, however, an equal
probability of overestimating the PSF anisotropy. Hence, the estimate
is a very conservative upper limit for the importance of
systematics. Such a systematic underestimate for PSF anisotropy would
give rise to very large ``E'' and ``B''-modes in the cosmic shear
measurements, which are not observed. Hence, in reality the change in
the ellipticity of the halo, caused by systematics, is much smaller than
our conservative estimate.

\end{document}